\documentclass[a4paper,fleqn]{cas-sc}
\usepackage[numbers]{natbib}
\def\tsc#1{\csdef{#1}{\textsc{\lowercase{#1}}\xspace}}
\tsc{WGM}
\tsc{QE}
\tsc{EP}
\tsc{PMS}
\tsc{BEC}
\tsc{DE}
\usepackage{braket}
\usepackage{bm}

\begin{document}
\let\WriteBookmarks\relax
\def\floatpagepagefraction{1}
\def\textpagefraction{.001}
\shorttitle{Optimization of wide-band quasi-omnidirectional 1-D photonic structures}
\shortauthors{V. Castillo-Gallardo et~al.}

\title [mode = title]{Optimization of wide-band quasi-omnidirectional 1-D photonic structures}

\author[1,2,3]{V. Castillo-Gallardo}[orcid=0000-0002-8157-3534]
\cormark[1]
\ead{ing.victorcg@gmail.com}
\credit{Conceptualization, Software \& Investigation, synthesis \& characterization, Writing}
\address[1]{Centro de Investigaci\'{o}n en Ingenier\'{i}a y
  Ciencias Aplicadas, Universidad del Estado de Morelos, Av.
  Universidad 1001, Col. Chamilpa, Cuernavaca, Morelos 62209,
  M\'{e}xico.}

\author[1,2,3]{Luis Eduardo Puente-D\'{i}az}[orcid=0000-0003-3719-8535]
\ead{fmatpuente@gmail.com}
\credit{Conceptualization, Software $\&$ Investigation}
\address[2]{Instituto de Ciencias F\'{i}sicas, Universidad
  Nacional Aut\'{o}noma de M\'{e}xico, Av. Universidad S/N,
  Col. Chamilpa, 62210 Cuernavaca, Morelos, M\'{e}xico.}
\address[3]{Facultad de Ciencias F\'{i}sico Matem\'{a}ticas,
  Universidad Michoacana de San Nicol\'{a}s de Hidalgo, Av.
  Francisco J. M\'ugica S/N 58030, Morelia, Mich., M\'{e}xico.}

\author[4]{D. Ariza-Flores}[orcid=0000-0002-1775-6598]
\ead{david1cool@gmail.com}
\credit{Investigation, Writing - review $\&$ editing}
\address[4]{CONACyT-Universidad Aut\'{o}noma de San Luis
  Potos\'{i}, Karakorum 1470, Lomas 4ta Secc, San Luis Potos\'{i},
  S.L.P., 78210, M\'{e}xico.}

\author[3]{H\'{e}ctor P\'{e}rez-Aguilar}[orcid=0000-0002-8572-1485]
\ead{hiperezag@yahoo.com}
\credit{Investigation and Supervision}

\author[2]{W. Luis Moch\'{a}n}[orcid=0000-0003-0418-5375]
\ead{mochan@fis.unam.mx}
\credit{Conceptualization, Funding acquisition, Investigation, Supervision, Writing - review $\&$ editing}

\author[1]{V Agarwal}[orcid=0000-0003-2168-853X]
\fnmark[1]
\ead{vagarwal@uaem.mx}
\credit{Conceptualization, Funding acquisition, Investigation, Methodology, Supervision, Writing - review $\&$ editing}

\cortext[cor1]{Corresponding author}
\fntext[fn1]{On sabatical leave at ICF-UNAM from CIICAP-UAEM}

\begin{abstract}
We have designed, optimized, fabricated and characterized highly 
reflective quasi-omnidirectional (angular range of $0-60^\circ$) 
multilayered structures with a wide spectral range. Two techniques, 
chirping (a continuous change in thicknesses) and stacking of 
Bragg-type sub-structures, have been used to enhance the 
reflectance with minimum thickness for a given pair of refractive 
indices. Numerical calculations were carried out employing the 
transfer matrix method and we optimized the design parameters to 
obtain maximal reflectance averaged over different spectral ranges 
for all angles. We fabricated some of the optimized structures with 
porous silicon dielectric multilayers with low refractive contrast 
and compared their measured  optical properties with the 
calculations. Two chirped structures with thicknesses 21.6 $\mu$m 
and 60.4 $\mu$m, resulting in quasi omnidirectional mirrors with 
bandwidths of 360 nm and 1800 nm, centered at 1160 nm and 1925 nm 
respectively have been shown. In addition, we fabricated a stacked 
sub-Bragg mirror structure  with a quasi omnidirectional bandwidth 
of 1800 nm (centered at 1850 nm) and a thickness of 41.5 $\mu$m, 
which is almost two third (in thickness) of the chirped structure. 
Thus, our techniques allowed us to obtain relatively thin 
quasi-omnidirectional mirrors with wide bands over different 
wavelength ranges. Our analysis techniques can be used for the 
optimization of the reflectance not only in multilayered PS 
systems  with different refractive index contrasts but also for 
systems with other types of materials with low refractive index 
contrast. The present study could also be useful for obtaining 
omnidirectional dielectric mirrors in large spectral regions using 
different materials,  flat focusing reflectors, thermal regulators, 
or, if defects are included, as filters or remote 
chemical/biosensors with a wide angular independent response.
\end{abstract}




\maketitle

\section{Introduction}

Photonic crystal (PC)-based structures have been extensively investigated due
to their light controlling properties \cite{Joannopoulos2008,Goyal2018}.
Photons entering a PC interact with its periodically varying dielectric constant and
their energies are consequently organized into photonic bands. In
analogy to electronic bands withiny a
crystal, propagation is forbidden if the energy lies
within certain regions known as photonic band gaps (PBG's). Thus, the
PC behaves as a mirror if illuminated by light with frequencies
within in a PBG. For several applications it is useful to fabricate
{\em ommnidirectional} mirrors with an absolute band gap for all angles of incidence.
The use of metallic
mirrors is limited due to their relatively large absorption at the
visible (Vis) and near infrared (NIR) frequencies.  An attractive alternative
consists of low-absorption dielectric
structures,  which may be designed for specific frequency
ranges. One-dimensional (1D) and two-dimensional (2D) PCs have
applications in optoelectronics, optical telecommunications and computing,
laser technology \cite{Lopez2003,Masaya2010}, and radiative cooling applications
\cite{Kumar2020}. The simplest 1D-PC is composed of a finite number
of periodically alternating
layers of high $n_{H}$ and low $n_{L}$ refractive indices, with
corresponding thicknesses $d_H$ and $d_L$ chosen so that their optical
thicknesses corresponds to a quarter of a wavelength
$n_{H}d_{H}=n_{L}d_{L}=\lambda_{0}/4$ at a nominal
wavelength $\lambda_{0}$. This yields a {\em Bragg mirror} (BM)
which might have a large PBG including an omnidirectional gap if the contrast $n_H/n_L$
and the number of periods are large enough. Mirrors with high
reflectance within a wider frequency range may be engineered by
using different nominal wavelengths for different layers.
One way to obtain such structures is to gradually change the width of
the layers as a function of their depth, producing
\textit{chirped}-type structures \cite{Zipock1997}.
Another alternative is to stack BMs, in such a way that their reflection bands partially
overlap each other at their edges, resulting in a wider high-reflection band \cite{Xifre2009}.

On the other hand, different fabrication techniques have been used to
obtain omnidirectional mirrors (ODM) that operate on the visible and/or
near infrared (Vis-NIR) range of the electromagnetic spectrum. For
example, Chen et al. \cite{Chen1999},
reported the fabrication of a structure made of six pairs of
TiO$_{2}$/SiO$_{2}$  layers with
an omnidirectional band (ODB) of about 70 nm in the NIR range, employing a sol-gel deposition
method. Park et al. \cite{Park2003}, used molecular beam epitaxy to
grow a stack of four pairs of GaAs/AlAs layers, followed by its conversion to
GaAs/Al$_{2}$O$_{3}$ by selective oxidation of the AlAs
layers, to obtain an ODB from 710 to 950 nm.
On the other hand, DeCorby et al. \cite{DeCorby2005}, fabricated mirrors by
coupling multiple layers of Ge$_{33}$As$_{12}$Se$_{55}$ chalcogenide glass and
polyamide-imide deposited by thermal evaporation and spin-casting
respectively, to obtain a 150 nm wide omnidirectional band centered at
1750 nm. Furthermore, Jena et al. \cite{Jena2019},
used sequential asymmetric bipolar pulsed DC magnetron sputtering for TiO$%
_{2}$ layers and radio frequency magnetron sputtering for SiO$_{2}$
layers to generate TiO$_{2}$/SiO$_{2}$ 1D-PC's and achieved an ODB from
592 to 668 nm. However, these techniques are expensive and they
require sophisticated equipment and long fabrication times. Use of
porous silicon (PS)
is an attractive alternative, since it can be easily manufactured by
electrochemical etching crystalline Si in an hydrofluoric
acid based electrolyte to obtain a sponge-like nanostructure composed
of Si and air. By modulating the supplied current and its application
time during the electrochemical reaction, it is possible to obtain
multilayered structures allowing for the fabrication ODM's \cite{Xifre2015,Pavesi2000}.
Although optical filters \cite{Estevez2009,Ariza2014} are the most common
application of PS PC's, they have also been widely used as chemical
sensors \cite{Giusseppe2011,Agarwal2018}, waveguides \cite{Hussel1997} and
for photoluminescence control \cite{Antunez2014}. Recently, the study of
chirped multilayer structures has
increased due to their possible application as flat focusing mirrors \cite{Wu2021,Kozar2017,Cheng2018}.
PS based dielectric optical filters, quasi-ODM's and ODM's have been extensively studied in different
regions of the electromagnetic spectrum, such as the ultraviolet (UV)
\cite{Jimenez2020}, visible \cite{Ariza2012}, and NIR \cite{Bruyant2003} regions.
However, to obtain the desirable high index of refraction
contrast, layers of high porosity have to be employed, which makes the
resulting structures relatively fragile. In addition, use of high
porosity makes the fabrication of large structures, with a
very large number of layers, unfeasible. For this reason, in this work
we study different strategies to produce multilayered mirrors
with high reflectance omnidirectional band
with relatively small thickness and refractive index contrast.

This paper is organized as follows. In Sec. \ref{s:theory} we develop the
methods used to design and calculate the reflectance spectra of the PS
multilayered structures such as chirped structures and mirror
stacks. In Sec. \ref{s:exp} we provide details about the
fabrication of these structures. In Sec. \ref{s:res} we present the numerical
and experimental results corresponding to our proposed structures and compare
them to previous reports. We obtained significantly large band-widths
with relatively thinner structures than reported previously.
Finally, we discuss our conclusions in Section \ref{s:conc}.

\section{Theory} \label{s:theory}

For the analysis of the propagation of electromagnetic fields through
multilayered systems it is usual to employ the transfer matrix
method \cite{PerezA2004,Pochi2005}. Assuming the system is 1D, with variations
only along the $z$ directions, and that the polarization of light is either
transverse electric (TE) or transverse magnetic (TM), the
transfer matrix $M\left(z_{2},z_{1}\right)$ is
a 2x2 matrix that relates the components of the electric $E_{\| }$
and magnetic  $H_{\|}$ fields parallel to the $xy$ plane
evaluated at $z_2$ to their value at $z_{1}$,
\begin{equation*}
        \left(
        \begin{array}{c}
                E_{\|} \\
                H_{\| }%
        \end{array}%
        \right) _{z_{2}}=M\left( z_{2},z_{1}\right) \left(
        \begin{array}{c}
                E_{\| } \\
                H_{\| }%
        \end{array}%
        \right) _{z_{1}}.
\end{equation*}%
Many equivalent formulations have been proposed to obtain and use $M$
\cite{Mochan1987,Ortiz2020,Chavez2020}. Here we use a recently developed formalism
for the numerically stable calculation of the reflectance of large
multilayer systems, as
proposed by Puente-D\'{i}az, et al. \cite{Puente2020}, summarized in the
supplementary information. The
explicit expressions for the optical coefficients are:
\begin{equation}
        r=\sigma \frac{Z_{0}M_{11}+M_{12}-Z_{0}Z_{s}M_{21}-Z_s
                M_{22}}{Z_{0}M_{11}-M_{12}-Z_{0}Z_{s}M_{21}+Z_{s}M_{22}},  \label{rBloch}
\end{equation}%
and
\begin{equation}
        t=\frac{2\zeta}{Z_{0}M_{11}-M_{12}-Z_{0}Z_{s}M_{21}+Z_{s}M_{22}},  \label{tBloch}
\end{equation}%
where $M_{ij}$ are the elements
of the transfer matrix which transfer
the fields from its surface interface with vacuum at $z_0$ towards the
substrate at $z_s$,  $Z_0$ is the surface impedance of vacuum and $Z_s$ is the
surface impedance of the substrate, and we define $r=F_r/F_i$ and $t=F_t/F_i$ in terms
of the reflected and transmitted fields, where we choose the $F$'s as electric
fields for (TE) polarization and magnetic fields
for transverse magnetic polarization (TM). Here, $\sigma=-1$ and
$\zeta= Z_s$ for TE polarization, while
$\sigma=1$ and $\zeta=Z_0$ for TM
polarization.
The reflectance is given by
$R=|r|^{2}$ and the
transmittance by $T=\beta \left\vert t\right\vert ^{2}$ with $\beta
=\text{Re}Z_{0}/\text{Re}Z_{s}$ for TE polarization and $\beta =\text{Re}Z_{s}/\text{Re}Z_{0}$ for
TM polarization. The complex refractive index of the porous
layers were obtained through the Bruggeman's effective medium theory, which
has been reported to adequately reproduce the  optical parameters of PS
\cite{Giusseppe2011,Pap2006,Estrada2018}.

\section{Experimental details}\label{s:exp}

Some of the proposed photonic structures were synthesized through anodic
etching of a (100) oriented, $p$-type Boron doped, crystalline Si wafer with
resistivity 0.002-0.005 $\Omega \cdot $cm, under galvanostatic conditions
\cite{Canham1990,Escorcia2007}. The electrochemical anodization process was
performed at room temperature, with an electrolyte of aqueous
hydrofluoric acid (HF) (48$\%$ of
wt) and ethanol (99.9$\%$ of wt) in 1:1 volumetric proportion, respectively.
However, as it is not desirable to use very high
porosity contrasts due to structure fragility and electrolyte
diffusivity problems \cite{Ariza2011}, the current
densities were chosen as 35 and 305 $\text{mA}/\text{cm}^{2}$,
with corresponding porosities of 51\% and 76\%, respectively.
These porosities were determined through previously obtained calibration
curves using a gravimetric technique \cite{Pap2006}. The
etching rate of PS was obtained by synthetizing single layers under
similar conditions and measuring their thicknesses through Scanning
Electron Microscopy (SEM). Absolute reflectivity measurements were
carried out with a Perkin Elmer Lambda 950 UV/Visible
spectrophotometer with a variable angle universal reflectance
accessory (URA) for different incident angles $\theta _{i}=10^{\circ
}$, $20^{\circ }$, $30^{\circ }$, $40^{\circ }$, $50^{\circ }$ and
$60^{\circ }$ using non-polarized light. The maximum and minimum
values of $\theta _{i}$ were constrained by the angular range of the
URA.

\section{Results and discussion}\label{s:res}

In this section we present optimized calculations of the reflectance
of different multilayered omnidirectional wide-band mirrors and we
compare the results with experimental spectra taken from
the corresponding samples at diferent angles of
incidence. Sec. \ref{ss:chirped} is devoted to chirped-type Bragg
mirrors while Sec. \ref{ss:stacked} is devoted to structures made of
stacked sub-mirrors.

We develop two techniques to design
multilayer photonic structures. In the first, the wavelength
($\lambda _{j}$) to which the $j$-th pair of layers is tuned is determined by
a given function of $j$. Several functions are proposed and their
parameters are optimized to maximize the reflectance $\braket{R}$ averaged over
a given spectral and angular range.
The second technique consisted in stacking
sub-mirrors tuned to different wavelengths, chosen so that their
spectral ranges overlap. For each sub-mirror, the spectral range was
obtained from the dispersion relation of a corresponding periodic
structure. The degree of overlap was optimized to maximize $\braket{R}$.
The optimizations were done employing the Minuit module  \cite{minuit}
of the Perl Data Language (PDL) \cite{PDL1997}.

\subsection{Chirped-type Bragg mirrors}\label{ss:chirped}

Here we study multilayered structures where the thicknesses for each
pair $j=0\ldots N_p-1$ of layers are tuned to a wavelength $\lambda_j$ which changes
gradually with the depth of the layer, according to
\begin{equation}
        \lambda_{j}=\lambda_{\min }+\left(\lambda_{\max }-\lambda_{\min}\right) f\left(\frac{j}{N_p-1}\right),\label{Dis}
\end{equation}%
where $\lambda_{\min }$ and $\lambda_{\max }$ are the minimum and
maximum {\em design} wavelengths respectively, $N_p$ is the number of
periods  in the structure, and $f(\xi)$ is a smooth function that
goes from 0 to 1 as its argument goes from the surface ($\xi=0$) to the
substrate ($\xi=1)$. We only consider increasing functions $f$ due to
the high absorption of PS in the ultraviolet region, which decreases in
the visible and becomes negligible in the near infrared, so, that the first
periods are tuned in the UV-Vis regions. We consider the
following classes of functions:
\begin{equation}
        f_{1}\left(\xi \right) =\xi^{\alpha },  \label{F1}
\end{equation}
\begin{equation}
        f_{2}\left(\xi \right) =\frac{1}{2}\left(\xi^{\alpha }+\xi^{\beta}\right)
        \label{F2}
\end{equation}
and%
\begin{equation}
        f_{3}\left(\xi\right) =A\xi^{\alpha }\left(1-\xi\right) +\xi^{\beta },  \label{F3}
\end{equation}
where $\alpha$, $\beta$ and $A$ are parameters to optimize in order
to maximize the average reflectance $\braket{R}$ over given spectral
and angular ranges. The class
$f_{1}$ (Eq. (\ref {F1})) with  $\alpha>0$ corresponds to simple
increasing profiles, $f_2$ is the arithmetic mean
of two functions of the type $f_{1}$ with different
powers $\alpha$ and $\beta$, and $f_{3}$ (Eq. (\ref{F3})) is
designed so that the first term dominates near the surface and the
second near the substrate, to allow different behaviors
at the edges of the spectrum.
The parameters
$\alpha$, $\beta$, $A$ and $N_{p}$ were optimized to maximize the
reflectance for non-polarized light $\braket{R}$ averaged in the
angular range of $0^\circ$ to $90^\circ$ and either in the
spectral region $\mathcal R_1$ from 350 to
1400 nm or $\mathcal R_2$ from 400 to 3000 nm. For $\mathcal R_1$ we took
$\lambda_{\min}$ as another parameter to be optimized while $\lambda_{\max}$
was fixed at 1400 nm. For $\mathcal R_2$ we fixed both
$\lambda_{\min}=850$ nm and $\lambda_{\min}=3000$ nm.
In the first case, we decided to optimize $\lambda_{min}$ to
obtain the best design in the high absorption zone of the PS.
Approximately $90\%$ of the solar radiation that reaches the
earth's surface is contained in the spectral region
$\mathcal R_1$, and $98\%$ in $\mathcal R_2$ \cite{Bird1983}.
A direct
application of the optimized structures in these regions is solar
reflectors. The
optimized parameters corresponding to the choice $p_l=0.51$ and
$p_h=0.76$ for the low and high porosities
are shown in Table \ref{tab:tableA}. In the supplementary 
information, we include the parameters used to calculate the 
reflectance of structures corresponding to the porosity $p_l=0.30$ 
and $p_h=0.76$. We also add the theoretical results obtained for 
$p_l=0.42$ and $p_h=0.76$.

\begin{table}
        \centering
        \label{tab:tableA}
        \addtolength{\tabcolsep}{-2pt}
        \begin{tabular}{rrrrrrrrrr}
                Class&$\alpha$&$\beta$&A&$\lambda_{min}(nm)$&$N_{p}$&$d(\mu\text{m})$       &ODW(nm)&ODC (nm)&$\braket{R}(\%)$\\
                \hline
                \hline
                $f_{1}$     &0.24    &---      &---     &250&222   &86.1    &300    &1150    &89\\
                $f_{2}$     &0.37    &1.06     &---     &400&102   &35.9    &220    &1110    &88\\
                $\bm{f_{3}}$&\bf 1.23&\bf 0.540&\bf 0.18&\bf 320&\bf 63&\bf 21.6&\bf 250&\bf 1125&\bf 89\\
                \hline
                $\bm{f_{1}}$&\bf 1.20&\bf ---  &\bf --- &\bf ---&\bf 90&\bf 60.4&\bf 2150&\bf 1925&\bf 92\\
                $f_{2}$     &1.08    &1.86     &---  &---  &105   &61.7    &2000    &2000    &89\\
                $f_{3}$     &1.01    &1.95     &0.20  &---  &102   &62.3    &2050    &1975    &90\\
        \end{tabular}
        \caption{Optimized parameters $\alpha$, $\beta$, $A$ and $N_p$ yielding the highest
                reflectance $\braket{R}$ averaged over the spectral ranges
                $\mathcal R_1$ (first block) and $\mathcal R_2$ (second block) and
                the angular range $0-90^\circ$ for the profile classes $f_1$, $f_2$ and
                $f_3$ (Eqs. (\ref{F1}) - (\ref{F3})). We include the thickness $d$
                of the structure and the omnidirectional spectral
                width and center (ODW and ODC).}
\end{table}

\begin{figure}
        \label{Fig1}
        \begin{center}
                \includegraphics[width=\textwidth]{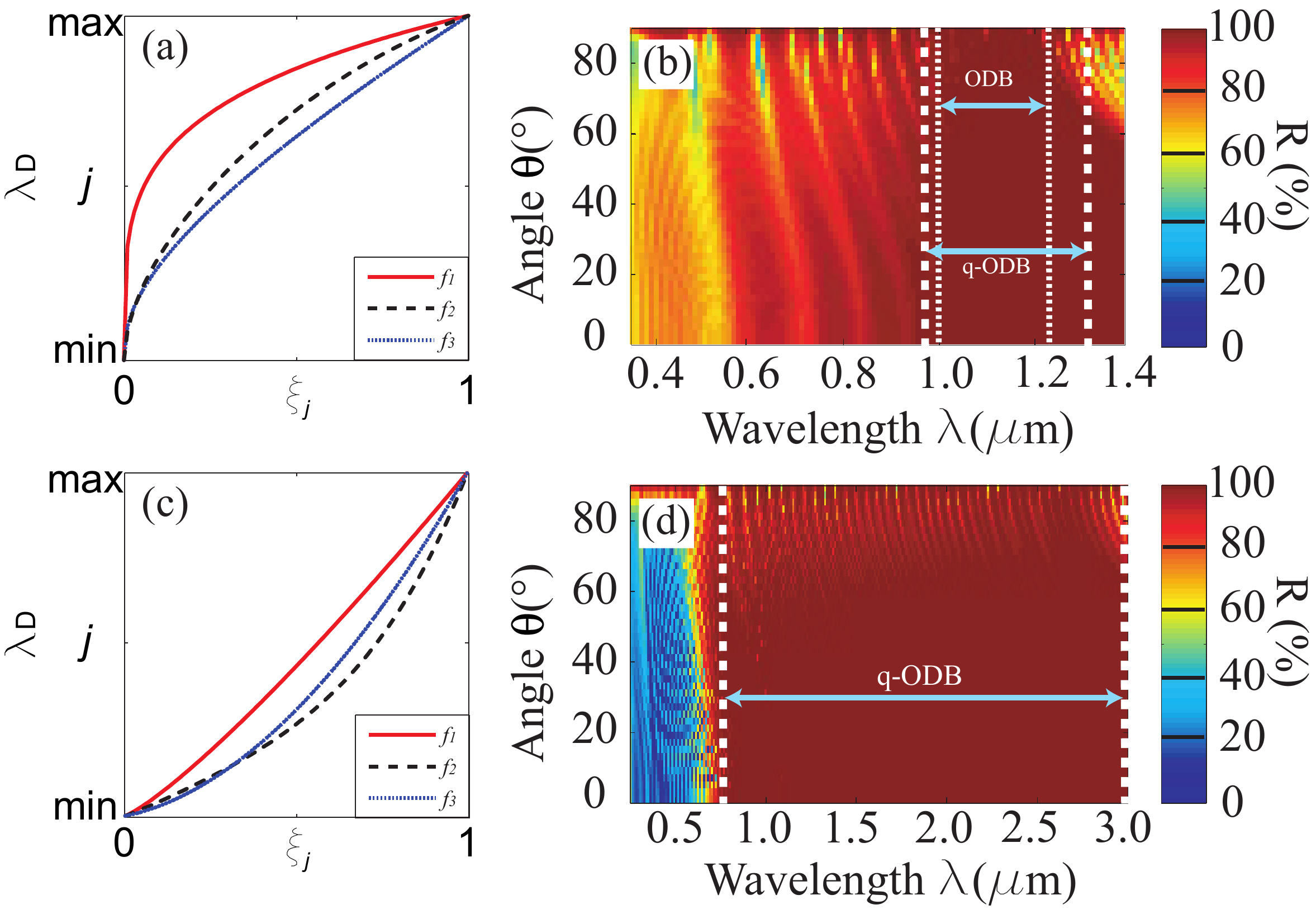}
        \end{center}
        \caption{Profiles of the functions $f_n(\xi_j)$ optimized for
                the region $\mathcal R_1$ from 350 to 1400 nm (panel a) and
                $\mathcal R_2$ from 400 to 3000 nm (panel c). Reflectance
                for non-polarized light
                vs. wavelength and angle of incidence of the structure
                optimized for the region $\mathcal R_1$ (panel b) and
                $\mathcal R_2$ (panel d). The dotted lines
                indicate the ODB and the dashed lines
                indicate the quasi-ODB.}
\end{figure}

The function $f_{3}$ yields after optimization the highest
$\braket{R}$ in the region $\mathcal R_1$ and with the thinnest
structure. Furthermore, this
structure has an omnidirectional width (ODW) of 250 nm
centered at (ODC) 1125 nm, as shown in Fig. \ref{Fig1}(b).  Various
photonic structures have been reported with
ODB's \cite{Xu2018,Ahmed2014,Reiner2000} defined as those spectral regions for
which $R>90\%$ for all angles. Here we we adopt the same criterium.
We synthesized a structure designated {\em sample $S_1$} corresponding
to these results. For the region $\mathcal R_2$ the
resulting structures have similar thicknesses and attain similar average reflectance,
though the one with the smallest thickness and the maximum $\braket{R}$
is that obtained from the family $f_{1}$. We
also synthesized the corresponding
structure and designated it {\em sample $S_2$}. The
calculated optimized reflectance spectrum is shown in Fig. \ref{Fig1}(d).
This structure has a very wide region where $R(\lambda, \theta)>
0.90$, which goes from 850 to 3000 nm, but in a restricted angular range
from $0^\circ$ to $70^\circ$. Thus, it has a
quasi-ODW of 2150 nm centered at 1925 nm.
The functions $f_n$ optimized for the region $\mathcal R_1$ are shown
in Fig. \ref{Fig1}(a), and those for $\mathcal R_2$
in Fig. \ref{Fig1}(c). The profiles $f_n$ are qualitatively
similar. For example, for $\mathcal R_1$ the optimal photonic
structures have a small number of periods tuned to the wavelengths where
silicon has a large absorption, i.e.,  the first periods of the
structure.
As withinin $\mathcal R_2$ the absorption of porous silicon is negligible, the optimized $f_n$ are not too
far from a linear profile.

\begin{figure}
  \label{Fig2}
  \begin{center}
    \includegraphics[width=\textwidth]{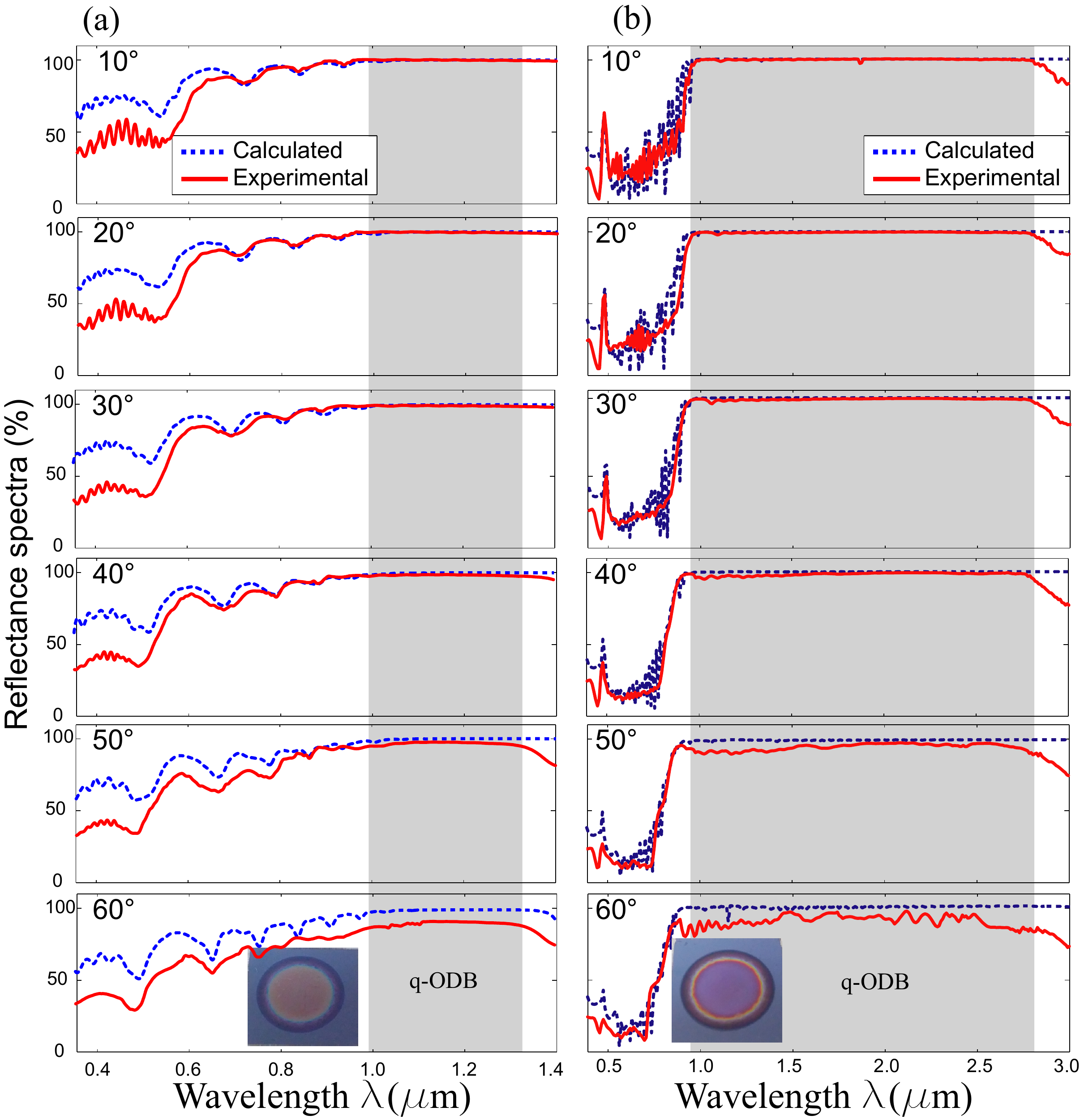}
  \end{center}
  \caption{Reflectance spectra for non-polarized light calculated and measured at different
    angles of incidence $\theta=10^\circ$, $20^\circ$, $30^\circ$,
    $40^\circ$, $50^\circ$, and $60^\circ$ for structure $S_1$
    (panel a)  and $S_2$ (panel b). The gray band indicates the
    region where the reflectance is greater than  $0.90$,
    corresponding to the quasi ODB of the
    structure. Inset: photograph of the synthesized structure.}
\end{figure}

In Fig. \ref{Fig2}(a) we present the calculated and measured
reflectance for non-polarized light corresponding to sample $S_1$ in the spectral range from 350 to 1400
nm for different angles of
incidence $\theta=10^\circ$, $20^\circ$, $30^\circ$, $40^\circ$,
$50^\circ$, and $60^\circ$. The spectra are qualitatively
similar, and for $\lambda>700$ nm and $10^\circ < \theta <
40^\circ$, they agree quantitatively, with a difference smaller
than $5\%$.
The
measured reflectance is lower than the calculated one, more so at
small wavelengths and large angles. This difference
may be partially attributed to the scattering of light due to the
roughness present at the sample
\cite{Theiss1994,Chavez2020,Ortiz2020} that we have not taken into
account in our theory. Due to the limitations of our
spectrophotometer, we couldn't perform measurements for $\theta>60^\circ$ and thus
check the ODB. Nevertheless, our results allow us to conclude that
the structure has a quasi-ODB from 980 and
1340 nm.
In Fig. \ref{Fig2}(b) we compare the
calculated and measured spectra of sample $S_2$ in the range
from 400 to 3000 nm.
Both spectra are very similar, with some differences at long
wavelengths which can be
attributed to an increase of imperfections such as wavy layers and
more roughness for deeper layers, due to a more restricted diffusivity
of the electrolyte as the thickness of the structure is increased
\cite{Negro2003,Vincent1993}. According to our results, this
structure has a wide quasi-ODW of 1800 nm
centered at 1880 nm.

\subsection{Photonic structure with sub-mirrors stacking}\label{ss:stacked}

Another class of techniques for designing good reflectors across a wide spectral range
is to stack sub-mirrors tuned to different wavelengths chosen with
different criteria.
For example, Agarwal et al. \cite{Agarwal2003} proposed a photonic
structure in which 54 sub-mirrors were stacked as follows: the
first sub-mirror consisted of 2.5 periods tuned to
$\lambda_{1}=700$ nm; the following $j$-th sub-mirrors were tuned to
wavelengths $\lambda_{j}$ according to the sequence $\lambda_{j+1}
= \lambda_{j}+ \left(2+j \right)$ nm
with 4 periods for
each of the following periods.
Furthermore, Estrada-Wiese et al. \cite{DelRio2018}
developed a method that combines three stochastic optimization algorithms together with
a  narrow space search methodology to obtain a customized and optimized
PC configuration.

We recall that for a periodic structure, the propagation of
electromagnetic waves is described by the dispersion relation
\begin{equation}
        \cos KD = \frac{1}{2} \text{Tr} \bm M,   \label{Dispersion}
\end{equation}
where $K$ is Bloch's vector and $\bm M$ the transfer matrix of a single
period of thickness $D$ \cite{Mochan1987, Mochan1988,
        Perez2018}. Thus,
whenever $\left\vert\text {Tr} \left(M \right) \right\vert>2$
electromagnetic energy cannot propagate in the periodic system. This condition corresponds to a
photonic band gap within which we
expect a high reflectance, even for a finite system made of only a few periods. This suggests
the following design strategy. We stack sub-mirrors
each made of a few periods of a periodic system such that its photonic band gap
(PBG) overlaps the PBG of the previous and that of the next
sub-mirror. The number of periods of each sub-mirror is an
important parameter. If it is too small, the sub-mirror would be
partially transparent within the PBG. If it is too large, the
resulting structure would be too thick. This is particularly important
for sub-mirrors tuned to shorter wavelengths, for which there is some
absorption. We followed thus two design
strategies: In the first, we considered sub-mirrors each with the same
number $p$ of periods. In the second, a different number of periods was
assigned to each sub-mirror according to the center of its PBG: we chose
a single period for those sub-mirrors $j$ tuned to
$\lambda_{j}<500$ nm, two periods for $500 <\lambda_ {j} < 650$ nm,
three periods for $650<\lambda_ {j}<800$ nm, and for
$\lambda_{j} \geq 800$ nm the number of periods $p$ was varied between
4 and 10. In both designs, the degree of overlap of consecutive PBGs was
taken as the parameter to be optimized in order to maximize
the reflectance $\braket{R}$ averaged within the spectral and
angular range $400\text{ nm}\le
\lambda\le 3000\text{ nm}$ and for $0^\circ\le\theta\le90^\circ$.
We define the percentage of overlap in terms of the portion of the PBG of the
$j$-th mirror that lies within the PBG of the $(j + 1)$-th
mirror.

Table \ref{tab:table2} shows the optimized values of the average
reflectance $\braket{R}$ for
different choices of the number of periods $p$ of each sub-mirror, as
discussed above, including the resulting number $N_m$ of stacked sub
mirrors, the thickness $d$ of the structure, and the optimal
overlap of consecutive PBGs. The first and second blocks correspond to
the first and second strategies discussed above.
\begin{table}[htbp]
        \centering
        \label{tab:table2}
        \begin{tabular}{rrrrr}
                $p$&$N_m$&$d (\mu\text{m})$&\% overlap&$\braket{R}$\\
                \hline
                \hline
                1&98&43.2&95&91.0\\
                3&31&42.5&81&91.8\\
                5&17&41.9&68&92.0\\
                \hline
                4&25&38.7&74&92.3\\
                \bf 6&\bf 18&\bf 41.5&\bf 78&\bf 95.4\\
                10&13&45.4&56&95.3\\
        \end{tabular}
        \caption{Total number $N_m$ of stacked sub-mirrors, thickness $d$,
                overlap between consecutive PBG and average reflectance
                $\braket{R}$ for optimized structures designed with a
                number $p$ of periods per sub-mirror (first block) or a number $p$
                of periods for those sub-mirrors tuned to wavelengths
                $\lambda_j\geq 800$ nm (second block, see text), for different
                values of $p$. The parameters corresponding to the best design
                are typeset in bold.}
\end{table}

\begin{figure}
        \begin{center}
                \includegraphics[width=\textwidth]
                {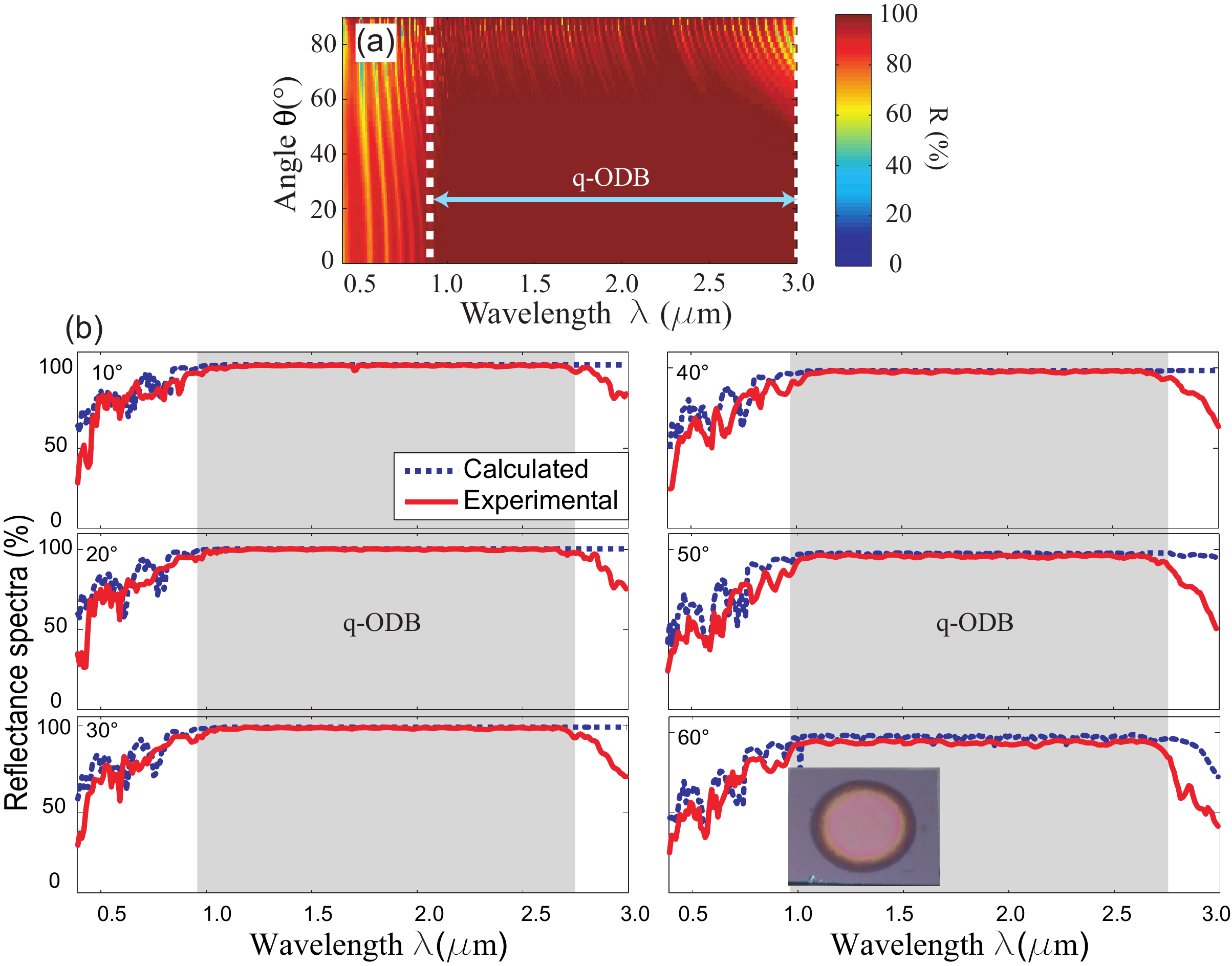}
        \end{center}
        \caption{ (a) Calculated reflectance $R(\lambda,\theta)$ for non-polarized light of the optimized structure $S_3$
                as a function of wavelength $\lambda$ and incidence angle
                $\theta$. (b) Measured and calculated reflectance spectra at several
                angles of incidence $\theta=10^\circ$, $20^\circ$,
                $30^\circ$, $40^\circ$, $50^\circ$, $60^\circ$. The gray band indicates
                the region in which the reflectance is greater than $R>0.90$.  Inset shows a
                photograph with a top view of the synthesized structure.}
        \label{Fig3}
\end{figure}

According to table \ref{tab:table2}, the structure that has the highest
reflectance is in the second block, with an optimal overlap between
consecutive PBGs of $78\%$, a thickness of
41.5 $\mu$m, and an average reflectance $\braket{0.954}$. It is made up
of 18 sub-mirrors tuned to the wavelengths
400, 460, 510, 570, 640, 720, 810, 910, 1020, 1150, 1290, 1450,
1630, 1830, 2060, 2320, 2610, and 2860 nm.
We synthesized this structure and designated it {\em sample $S_3$}.
Its calculated and measured
reflectance spectra are shown in Fig. \ref{Fig3}.

Fig. \ref{Fig3}(a) reveals a quasi-ODM
in the range from 950 to 2900 nm, with an angular range between $0^\circ$ and
$60^\circ$. It is also observed that at incidence angles $\theta
>70^\circ$ and long
wavelengths the reflectance decreases and has oscillations, which we attribute to the
greater penetration of the electromagnetic field within the structure, as shown by
Puente-D\'{\i}az, et al. \cite{Puente2020}. In Fig. \ref{Fig3}(b) we compare the
calculated and measured reflectance for the structure $S_3$ at several angles of
incidence. Although, the quasi omnidirectional wavelength range of
the synthesized structure, from 950 to 2750 nm, is slightly smaller than the calculated
one, the measured PBG is almost twice as wide
as a recently reported PS chirped structure \cite{Chavez2020}, which
has the largest previously reported ODW.

Finally, Table \ref{tab:table3} shows data from some porous
silicon based ODM designed to operate in the NIR
region and their of the ODB's are compared with those of
the structures analyzed in the present study. The ODW
of the structure $S_1$ results 20 nm wider than that reported by Bruyant,
et al. \cite{Bruyant2003} and it is located closer to the visible range. Furthermore,
the ODB of the structure $S_3$ begins at 950 nm like the one
synthesized by Estevez, et al. \cite{Estevez2009}; nevertheless, its width extends to 2750
nm. In other words, $S_3$ has an ODW 3.55 times higher. Moreover, the
ODW of the structure reported by Xifre-Perez, et al. \cite{Xifre2005} represents only 17.6
\% of the ODW of our $S_2$ and $S_3$ structures. Aditionally, the table
reveals that the porous silicon based
structures $S_2$ and $S_3$ have the widest gap widths, 1.36 times greater than the
largest reported work until now in the NIR region \cite{Wu2021}. 
As compared to the previous studies \cite{Bruyant2003,Estevez2009,Xifre2005,Wu2021,Fink1998}, this work 
opens the possibility of generating dielectric multilayered 
structure with large omnidirectional band with 
the materials with relatively low refractive index contrast.

\begin{table}
        \centering
        \label{tab:table3}
        \begin{tabular}{rrrrr}
                Reference&Year&Sample&ODB (nm)&ODW (nm)\\
                \hline
                \hline
                Bruyant, et al. \cite{Bruyant2003}      &2003&-    &1100-1440&340\\
                Xifre-P\'{e}rez, et al. \cite{Xifre2005}&2004&-    &1297-1615&318\\
                Estevez, et al. \cite{Estevez2009}      &2009&-    &950-1456 &506\\
                Chavez, et al. \cite{Chavez2020}        &2020&-    &1000-2000&1000\\
                Wu, et al. \cite{Wu2021}                &2021&-    &1284-2605&1321\\
                \hline
                &    &$S_1$&980-1340&360 \\
                This work                               &2021&$S_2$&980-2780&1800\\
                &    &$S_3$&950-2750&1800\\
       
        \end{tabular}
        \caption{Development of porous silicon based omnidirectional mirrors for the NIR region for
                previously published works (first block) and for the different samples studied here (second
                block. We show the attained omnidirectional bands (ODB) and their widths (ODW).}
\end{table}

\section{Conclusion}\label{s:conc}

We have designed, optimized, fabricated and characterized highly reflective quasi
omnidirectional (angular range of $0-60^\circ$) multilayered structures with a wide
spectral range. Two techniques, chirping (a continuous change in thicknesses) and
stacking of Bragg-type sub-structures, have been used to enhance the reflectance with minimum thickness for a given pair of refractive indices. Numerical calculations were
carried out employing the transfer matrix method and we optimized the
design parameters to obtain maximal reflectance averaged over different spectral ranges for
all angles. 
We fabricated some of the optimized structures with
porous silicon
dielectric multilayers with low refractive contrast and compared their measured  optical properties
with the calculations. Two chirped structures
with thicknesses 21.6 $\mu$m and 60.4 $\mu$m,
resulting in quasi omnidirectional mirrors with bandwidths of 360 nm and 1800 nm, centered
at 1160 nm and 1925 nm respectively have been shown. In addition, 
we fabricated a stacked sub-Bragg mirror structure  with a quasi 
omnidirectional bandwidth of 1800 nm (centered at 1850 nm) and a 
thickness of 41.5 $\mu$m, which is almost two third (in thickness) 
of the chirped structure.
Thus, our techniques allowed us to obtain relatively thin quasi-omnidirectional mirrors with wide bands
over different wavelength ranges.
Our analysis techniques can
be used for the optimization of the reflectance not only in multilayered PS systems with
different refractive index contrasts but also for systems with other types of
materials with low refractive index contrast. The present study could also be useful for obtaining omnidirectional dielectric mirrors in large spectral regions using different materials,  flat focusing reflectors, thermal regulators, or, if
defects are included, as filters or remote chemical/biosensors with 
a wide angular independent response.

\section*{Acknowledgments}

This project is supported by Consejo Nacional de Ciencia y
Tecnolog\'{i}a through grant No.A-S1-30393(VA), 256243 (DA), C\'{a}tedras
Conacyt program 1577 (DA), and by DGAPA-PAPIIT UNAM under grant IN111119 (LM).
VA acknowledges PRODEP (Sabbatical grant), and
is grateful to ICF where she spent the sabbatical year.
HPA also expresses his gratitude to the Coordinaci\'{o}n
de la Investigaci\'{o}n Cient\'{i}fica de la Universidad
Michoacana de San Nicol\'{a}s de Hidalgo. VCG is grateful to CIICAp
and ICF where he did this work.

\printcredits


\end{document}